\documentstyle[12pt]{article}
\begin{document}

\begin{titlepage}  
\thispagestyle{empty}
\title{ On the Evolution of a Large Class of Inhomogeneous Scalar Field  
Cosmologies  }  
  
\author{ J. Ib\'a\~{n}ez and I. Olasagasti  \\  
Dpto. F\'{\i}sica Te\'orica, Universidad del Pa\'{\i}s Vasco \\
Apdo 644 
48080 Bilbao, 
Spain. }  
\maketitle  
\vskip 2cm  
  
\newpage
\begin{abstract}  
The asymptotic behaviour of a family of inhomogeneous scalar field
cosmologies with exponential potential $V=\Lambda e^{k\phi}$ is studied.
By introducing new variables we can perform an almost complete analysis of
the evolution of these cosmologies. Unlike the homogeneous case
(Bianchi type solutions), when $k^2<2$ the models do not isotropize due to
the presence of the inhomogeneities.
\end{abstract}  
  
\end{titlepage}  
  
\section{Introduction}  

In the last years there has been a renewed interest in studying scalar
field cosmological solutions mainly due to the prominent role that
scalar fields would play in the early stages of the universe driving a
period of inflationary expansion. Scalar fields can be thought as
arising from the the low-energy limit of string theory or as coming from
the dimensional reduction in Kaluza-Klein models. 

One of the most interesting questions that has been analysed is the
so-called cosmic no-hair conjecture which states that the dynamics of a
scalar field with a realistic potential could isotropize the space time
smoothing out the inital anisotropies and inhomogeneities, giving at
late times a FRW universe. Initially, the cosmic no-hair conjecture was
proved for space times with a cosmological constant \cite{wald}. Since
during the ``slow-rolling" epoch the scalar field behaves as a
cosmological constant, that result was supposed to be applicable to
scalar field cosmologies. However, it was soon realised that the picture
would change when the dynamics of the scalar field is taken into account
\cite{heu}. The study of the evolution of cosmological models with
scalar field has been done mainly for homogenous solutions (Bianchi
type) and, in particular, when the potential of the field is of
Liouville type (exponential potential) \cite{bian}. Exponential
potentials arise in dimensional reduction theories and it is well known
that they lead to power law inflation in FRW models. 

Although there have been several attempts to study the evolution of the
scalar field in inhomogeneous models \cite{sak}, there is a lack of
results concerning how spatial inhomogeneities affect the dynamics of
the scalar field. It was shown for a particular family of exact
inhomogeneous solutions with exponential potential that there are cases
for which the scalar field does not guarantee that the models inflate or
isotropize \cite{fi}. Since a generic relativistic solution near the
initial singularity is neither homogeneous nor isotropic it is worth
analysing the effect of spatial inhomogeneities on the isotropization. 

In a previous paper (hereafter Paper I) \cite{ji} we started the study
of the asymptotic behaviour of a particular class of inhomogeneous
metrics described by a diagonal $G_2$ line-element. In that work we
analysed a particular family of $G_2$ self-similar solutions with
exponential potential. The main conclusions were that the asymptotic
behaviour of the model does not depend on the parameters describing the
spatial inhomogenity but rather on the parameter $k$, associated with
the slope of the exponential potential and the behaviour, essentially,
was the same as that found for homogeneous solutions, i.e: when $k^2<2$
the models isotropized. It was surmised that these conclusions could be
explained in terms of the ``weak" spatial inhomogeneity and that the
introduction of more ``strong'' inhomogeneities could result in a
different behaviour with respect to the homogeneous case. Since our
analysis is based only on the study of the equilibrium points, the
conclusions we draw have to be considered at the conjecture level, until
a more complete analysis of the structure of the field equations for
$G_2$ metrics is done. 

In this paper we extend the analysis of Paper I to a more general class
of $G_2$ exponential potential cosmologies and we show how in those
cases for which the homogeneous solutions isotropized (when $k^2<2$),
here the inhomogeneities prevent the isotropization. Since our analysis 
is based only on the study of the equilibrium points, the conclusions we
draw have to be considered at the conjecture level, until a more
complete analysis of the structure of the field equations for $G_2$
metrics is achieved.

It is known that scalar field homogeneous cosmological solutions with an
exponential potential evolve towards self-similar solutions and it was
conjectured that the same is true for $G_2$ solutions \cite{w1}. In this
paper we prove that the asymptotic metrics of the class of solutions
studied here are self-similar. 

The plan of the paper is the following: in Section 2 we will describe
the metric and the assumptions we make in order to simplify the
equations. A new set of variables is introduced in such a way that the
phase-space of the system is bounded for most of the cases considered.
In Section 3 the equilibrium points of the system and the correponding
solutions are studied. Finally the conclusions are explained in Section
4.

\section{The metric and the dynamical systems}  
In this paper we shall be considering spatially inhomogeneous
$G_{2}$ cosmologies admitting a two-parameter abelian
group of isometries for which the line element is written as:

\begin{equation}   
ds^{2}=e^{F} \left(-dt^{2}+dz^{2} \right)+G \left( e^{p} dx^{2}+   
       e^{-p} dy^{2} \right),  
\label{mod} 
\end{equation}  
with all functions appearing in the metric depending on both variables   
$t$ and $z$. In paper I we considered that the transitivity surface 
function $G$ was a homogeneous function, depending only on $t$,
and that the rest of the metric
functions were separable.
In the present case we would like to extend that 
analysis allowing the function $G$  
to depend on the variable $z$ as well, keeping the separability of the
metric functions.

The local behaviour of the models described by metrics of the form
(\ref{mod}) is determined by the gradient of the so-called transitivity
surface function $G$ \cite{carm}. A globally null or spacelike gradient
of this function corresponds to plane or cylindrical gravitational
waves. If the gradient is globally timelike, however, the metric
describes a cosmological model with spacelike singularities and if the
character of the gradient changes throughout space time it describes
colliding gravitational waves or cosmologies with timelike and spacelike
singularities. 

As the matter source we shall consider a minimally coupled scalar field
with exponential potential: 
  
\begin{equation}   
V( {\phi} ) = \Lambda e^{k {\phi}}.      
\quad   
\Lambda \geq 0   
\end{equation}   
For vanishing $k$ this will be equivalent to a cosmological constant 
term.  
The corresponding stress-energy tensor will read:  
  
\begin{equation}   
T_{ab}={\phi}_{,a} {\phi}_{,b}-g_{ab} \left[ \frac{1}{2}    
{\phi}_{,c} {\phi}^{,c} + V( {\phi}) \right].   
\end{equation}
It is well known \cite{tt} that this stress-energy tensor is
equivalent to a perfect fluid as long as the  
gradient of the scalar field is timelike.

For the model under consideration the resulting Einstein and
Klein-Gordon field equations are:
  
\begin{equation}  
\ddot{\phi} - {\phi}'' + {\dot G\over G} \dot{\phi} - {G'\over G} {\phi}' +  
e^{\displaystyle F} {\partial V \over \partial {\phi}}= 0,\label{KGE}  
\end{equation}  
\begin{equation}  
{\ddot G\over G} - {G''\over G} = 2\,e^{\displaystyle F}\,V({\phi}), 
\label{eq1} 
\end{equation}  
\begin{equation}  
\ddot p - p'' + {\dot G\over G} \dot p - {G'\over G} p' = 0,  
\end{equation}  
\begin{equation}  
{\dot G'\over G} - {1\over 2} {\dot GG'\over G^2} + {1\over 2} \dot pp' -  
{1\over 2} F'{\dot G\over G} - {1\over 2}\dot F{G'\over G}=-
\dot{\phi}{\phi}',  
\label{eq2}
\end{equation}  
\begin{equation}  
{\ddot G\over G}+{G''\over G}-{1\over 2}\left({\dot G\over G}\right)^2  
-{1\over 2}\left({G'\over G}\right)^2-\dot F{\dot G\over G}-F'{G'\over G}  
+{1\over 2}\dot p^2+{1\over 2}p'^2=-\dot{\phi}^2-{\phi}'^2,  
\end{equation}  
\begin{equation}  
{\ddot F}-F''-{1\over 2}\left({\dot G\over G}\right)^2  
+{1\over 2}\left({G'\over G}\right)^2  
+{1\over 2}\dot p^2-{1\over 2}p'^2=-\dot{\phi}^2+{\phi}'^2.  
\end{equation}

We want to find the general behaviour of the models under consideration
and, in particular, their behaviour near the initial singularity and at
late times. The treatment of the asymptotic behaviour of the
inhomogeneous metrics was initiated in \cite{w1} by extending the method
developed for homogeneous Bianchi type solutions with a perfect fluid
\cite{w2}. This method is based on the introduction of a new set of
adimensional variables constructed from the fluid quantities. For
inhomogeneous metrics the new variables satisfy a system of partial
differential equations and the equilibrium points are considered as
dynamical equilibrium points. Since the source of our metrics, the
scalar field, is not globally equivalent to a perfect fluid we can not
apply this method and we have to tackle the problem of resolving the
asymptotic behaviour of system (4)-(9) in a different way. In paper I we
analised a particular family of solutions of the above system. In
particular, we considered a family of solutions for which the
inhomogeneity was linear and introduced new dynamical variables in such
a way that the phase-space was compact and, therefore, we could describe
completely the asymptotic behaviour of the system. We will follow the
same approach as that already used in paper I. By fixing the spatial
dependence of the metric functions we will write this system of partial
differential equations as a set of ODE's and we will use standard
dynamical system techniques to find out the structure of the
phase-space. 

To obtain the desired set of ODE´s we have made a number of simplifying
assumptions on the metric functions: 

i) As in \cite{ajp} we write the scalar field $\phi$ in terms of a new
function $\psi$ as:

\begin{equation}  
{\phi}=-\frac{k}{2} \log(G)+ \psi . 
\label{assu1}  
\end{equation}  
Substituting (10) into the Klein-Gordon equation (4) we find that $\psi$
satisfies the same equation as function $p$. This allows us to choose
the function $\psi$ proportional to $p$: 
  
\begin{equation}  
\psi (t,z)=m\,p(t,z)   
\end{equation}  
where $m$ is an arbitrary constant.

ii) to simplify the equations further we consider separability of the
metric functions and in particular we assume the {\em ansatz}, already
used by different authors \cite{seno},\cite{ajp}: 
  
\begin{eqnarray}  
  &  G(t,z)=T(t)Z(z) & \\  
  &  F(t,z)=f(t)+f_{1}(z) & \\  
  &  e^{p(t,z)}=Q(t){Z(z)}^{n}. &    
    \label{assu2}    
\end{eqnarray}  
  
These are {\em ad hoc} assumptions, made with the only motivation of
simplifying the Einstein equations. They have been suggested by previous
works on inhomogeneous metrics with perfect fluid source. 

With these assumptions, the two functions depending on the spatial
variable $z$, that is, $f_{1}(z)$ and $Z(z)$ are inmediately fixed by
the equations. In fact, we see that Eq(6) separates into: 
  
\begin{equation}  
n \frac{Z''}{Z}=\frac{\dot{T}}{T} \frac{\dot{Q}}{Q} + \frac{\ddot{Q}}{Q}  
  - \left( \frac{\dot{Q}}{Q} \right) ^{2}=n \epsilon a^{2}, \;\; \; 
\epsilon=(-1,0,1),  
\end{equation}  
where $n\epsilon a^{2}$ stands for an arbitrary separation constant.   
Resolving the above equation we get:   
  
\begin{equation}  
Z(z)=\left\{ \begin{array}{ll}  
            A \cosh (az)+B \sinh (az)  &  \epsilon=1  \\  
            Az+B  &  \epsilon=0  \\  
            A \cos (az)+ B \sin (az) &  \epsilon=-1  
            \end{array}  
      \right.  
\end{equation}  
where $A$ and $B$ are arbitrary constants of integration.   

From Eq.(\ref{eq1})  when $\Lambda \neq 0$  
or from Eq(\ref{eq2}) when $\Lambda=0$ we find that:  
\begin{equation}
f_{1}(z)=2\alpha \log Z(z)+ \log D .  
\end{equation}  
When $\Lambda=0$, $\alpha$ and $D$ are arbitrary constants, but when
$\Lambda \neq 0$, $D$ is still an arbitrary constant but 
\begin{equation}
2\alpha=C \equiv k^2/2-kmn.
\end{equation}

The rest of the equations separate in a straightforward way giving rise
to new separable constants and a few relations betwen them. We shall not
give here the details of the calculations but only summarize them. 

When $Z(z) \neq e^{\pm az}$ we obtain from Eq.(8) or from Eq.(9) the
following constraint on the parameters: 
 
\begin{equation} 
 2\alpha=n^2M^2+C-\left({1 \over 2}+{k^2 \over 4} \right), 
\end{equation}  
where we have defined $M^2 \equiv 1/2+m^2$.  
When $\Lambda =0$ the above equation fixes the value of $\alpha$, but 
when $\Lambda \neq 0$ it becomes, from (18),
\begin{equation}  
n^2M^2={1 \over 2}+{k^2 \over 4}.  
\end{equation} 
The last relation allows us to write $m$ in terms of $n$ and $k$. The 
two solutions are:  
\begin{equation}  
m_{\pm}=\pm {\sqrt{k^2+2-2n^2} \over 2n}   
\label{m}
\end{equation}
which also gives $C$:  
\begin{equation}  
 C_{\pm}={k \over 2}(k \mp \sqrt{k^2+2-2n^2}).   
\end{equation}  
  
From now on we assume that Eqs(19)-(22) hold. That means that, in the
exponential case, we are only considering those solutions for which the
constants verify the above relations. In the general exponential case
all the solutions are self-similar and the constant $m$ is not
restricted by relation (\ref{m}). Since this case will not add any new
insights to the main conclusions of the paper we leave the corresponding
analysis to be discussed elsewhere. We are also assuming that $\Lambda
\neq 0$. Solutions with $\Lambda=0$ shall be limiting cases from which
all the solutions with $\Lambda\ne 0$ start their evolution. 

Finally the line element and scalar  
field have  the form:  
\begin{equation}  
ds^2=De^{f}Z^{2\alpha}(-dt^2+dz^2)+TQZ^{(1+n)}dx^2+{T \over 
Q}Z^{(1-n)}dy^2
\end{equation}
\begin{equation}  
 \phi={-k \over 2} \log T+m \log Q +(mn-k/2)\log Z.   
\end{equation}  
Where the function $Z(z)$ is defined in (16).  

Among the solutions described by the line element (23) we can identify
the following known solutions, all of them correponding to the case when
$\alpha=0$: If $Z(z)=\sin(az)$, or $\cos(az)$, $De^{f}=QT$ and $n^2=1$,
then the solutions reduce to the Kantowski-Sachs models; if $Z(z)=e^{\pm
az}, $ then one obtains that: when $n^2=0 $ (23) reduces to a type
Bianchi V solution, when $n^2=1$ to a type Bianchi III and when $n^2
\neq 0,1 $ to type Bianchi VI. Finally, if $Z(z)=$constant (23) is a
type Bianchi I solution. 

We can see also from Eq.(24) that when $mn=k/2$ the scalar field is
homogeneous and, therefore, can be cast in a perfect fluid form. In this
case the metric (23) belongs to the classes of solutions studied by
different authors \cite{ajp}, \cite{seno}. 

Another important class of solutions belonging to the line element (23)
is that obtained when $k=m=0$ (from Eq.(20) $n^2=1$). In this case the
scalar field vanishes and the source of the metric is a cosmological
constant. 

Once the spatial dependence has been fixed, we have to find the time
dependence of the metric that is determined by the three independent
functions $T(t)$, $Q(t)$ and $f(t)$. We introduce, for later
convenience, a new function $h(t)$ defined as: 

\begin{equation}  
h(t)=f+km \log Q .
\end{equation}
   
The Einstein equations, with the above assumptions, reduce to:

\begin{equation}
\frac{\ddot{T}}{T}=\epsilon a^{2}  
       +2\Lambda D \exp \left( h-\frac{k^{2}}{2} \log T \right), 
\label{self1}  
\end{equation}

\begin{equation}
\frac{\ddot{Q}}{Q}=n \epsilon a^{2}-\frac{\dot{T}}{T} 
\frac{\dot{Q}}{Q}  
       + \left( \frac{\dot{Q}}{Q} \right)^{2},
\end{equation}

\begin{equation} 
\ddot{h}={k^2\over 2}\epsilon a^{2} 
         -M^{2} \left( \frac{\dot Q}{Q} \right)^{2}  
         -\left( \frac{k^{2}}{4}-\frac{1}{2} \right)   
          \left( \frac{\dot T}{T} \right)^{2},  
\end{equation}

\begin{equation}  
 \frac{\dot T}{T}  
          +2n M^{2}  \frac{\dot{Q}}{Q} - \dot h=0, 
\end{equation}

\begin{equation}
\dot h  \frac{\dot T}{T} - M^{2} \left( \frac{\dot Q}{Q} \right)^{2}  
         -\left( \frac{k^{2}}{4}-\frac{1}{2} \right)   
          \left( \frac{\dot T}{T} \right) ^{2} =  
          2\Lambda D \exp (h- \frac{k^{2}}{2} \log T) +2 \epsilon a^{2}. 
\label{lig2} 
\end{equation}

From the original system of PDE's we have obtained a system of five
ODE's. These equations separate into three dynamical equations and two
constraint equations. Since equation (\ref{self1}), however, depends on
$T$ and its second derivative the associated dynamical system will have
four first order equations and two constraint equations. On the other
hand, the constraint (\ref{lig2}) will serve to eliminate the parameter
$\epsilon a^2$ from the system. 

To put this system of equations in the usual dynamical system form, we
make an appropriate change of variables. We want the new variables to be
normalized in such a way that they remain bounded near the initial
singularity. 

From Eq.(\ref{lig2}) we see that an appropriate set of new variables
$\beta$, $\Phi$ and $\Gamma$ is:

\begin{equation} 
\beta=\frac{M \frac{\dot{Q}}{Q}}{b \frac{\dot T}{T} +d \dot h}  
\makebox[0.7cm]{}  
\Phi=\frac{d\, \dot h}{b \frac{\dot T}{T} +d \dot h}
\makebox[0.7cm]{}   
\Gamma=\sqrt{2\Lambda D}\;\;{ \exp \left({1\over 2}\left[h-\frac{k^{2}}{2}
 \log T\right]\right) 
\over b \frac{\dot T}{T} +d \dot h} 
\label{nv}
\end{equation}  
with
\begin{equation}
\begin{array}{lll}
b^{2}= \frac{\mid 2-k^{2} \mid}{4}, &  d^{2}={1 \over \mid 2-k^{2} \mid}
& k^{2} \neq 2  \\  
b=d=1   &  &   k^{2}=2
\end{array}
\end{equation}

$\Gamma$ is related with the potential of the scalar field, it vanishes
when the scalar field is massless, $\beta$ is related with the time
dependent part of the function $p$ in the metric (\ref{mod}) and $\Phi$
depends on $f$. Unlike the Bianchi type solutions, where the kinematical
quantities of the source are used as variables, here, since the scalar
field is not globally timelike, we can not use that approach. The reason
for defining the variables as in (\ref{nv}) is to get a bounded phase
space (at least for a large subset of the solutions). Thus there is no
particular physical meaning for these variables, although another set of
variables with some physical meaning could perhaps be found. 

In terms of the new variables (31) and substituting $\epsilon a^2$ with
the aid of Eq.(30), Eq.(26) becomes:
\begin{equation}
\frac{\ddot T}{T}=\left[ \frac{1}{2}b^2
\frac{\Gamma^2-\beta^2}{\left( 1-\Phi\right)^2}+
\frac{1}{2}\frac{b}{d}\frac{\Phi}{1-\Phi}-\frac{1}{8} 
\left(k^2-2\right)\right]
\left(\frac{\dot T}{T}\right)^2.
\end{equation}
To decouple this equation from the rest of equations of the system 
we define a new time $\tau$, through the following expression:
\begin{equation}
 \dot \tau=b \frac{\dot T}{T} +d \dot h=\frac{b}{1-\Phi}\;\frac{\dot T}{T}.
\end{equation}

From the general behaviour near the initial singularity found by
Belinskii {\em et al} \cite{b}, we can assume that our models, for
constant $z$, behave near the initial singularity like a Kasner model,
in the sense that when $t \rightarrow 0$ 
\begin{equation}
 T \sim t, \makebox[1cm]{}  f\sim \log Q\sim \phi \sim \log t .
\end{equation}
This means that $\dot \tau \sim 1/t$ when $t \rightarrow 0$ which
guarantees the regular behaviour of the new variables near the initial
singularity. To see how $\tau$ behaves far from the singularity we  
write Eq.(\ref{lig2}) in the following way:

\begin{eqnarray}
\dot\tau^2 & = &
2\Lambda D\exp (h- \frac{k^{2}}{2} \log T) +2 \epsilon a^{2}+ M^{2}
\left( \frac{\dot Q}{Q} \right)^{2} \nonumber \\
 & & \mbox{} + d^2{\dot h}^2 
+\left( b^2+{k^2-2\over 4}\right)  \left( \frac{\dot T}{T} \right) ^{2}
\end{eqnarray}
when $k^2\neq 2$ and
\begin{equation}
\dot\tau^2 = 
4\Lambda D\exp (h- \log T) +4 \epsilon a^{2}+ 2M^{2} \left( \frac{\dot Q}{Q} 
\right)^{2}+
 {\dot h}^2 
+  \left( \frac{\dot T}{T} \right) ^{2}
\end{equation}
when $k^2=2$.

As long as $\epsilon=0,1$ all the terms in the right hand side of the
above two expressions are positive and, therefore, $\dot\tau$ is
different from zero for all values of $t$. Because of this we can
restrict our attention to $\dot \tau >0$ which implies that our new time
$\tau$ will grow monotonically with $t$. Since $\dot \tau \sim 1/t$ for
$t \sim 0$, the range of $\tau$ goes from $- \infty$ to $+ \infty$. On
the other hand, the situation with $\epsilon=-1$ is completely
different. Since there is nothing preventing $\dot \tau$ from being zero
at some finite time $t$, there will be solutions whose trajectories in
the phase-space of the new variables will escape to infinity in a finite
time. That means that, in this case, $\tau$ will no longer serve to
monitor the evolution in the coordinate time $t$. 

We are now in a position to write the equations in terms of these
variables. Since the change of variables does not depend on $k$ in a
continuous way, the system of equations does not have the same form for
all the values of $k$. First we write the constraint equations (29) and
(30) that in these new variables take particularly simple expressions.
The first constraint equation (29) relates $\Phi$ and $\beta$: 
\begin{eqnarray}   
\Phi & = & \frac{2+\sqrt{4-k^{4}}\: \beta}{4-k^{2}}\qquad  k^{2}<2  
\nonumber \\   
\Phi & = & \beta +\frac{1}{2}\qquad k^{2}=2 \\  
\Phi & = & \frac{2+\sqrt{k^{4}-4}\: \beta}{k^{2}} \qquad k^{2}>2. 
\nonumber  
\end{eqnarray}
As for the second constraint Eq.(30) 
it takes the simple form:  
  
\begin{equation}  
L(\beta,\Phi,\Gamma)\equiv  1- \beta^2- \Phi^2 
-\Gamma^2+H(\beta,\Phi,\Gamma)=\frac{2 \epsilon a^{2}}{{\dot \tau}^{2}}   
\end{equation}  
where
\begin{equation}  
\left\{  
\begin{array}{ll}  
                H=0 & k^{2}<2 \\ *[8pt]  
                H=\Phi-1 & k^{2}=2  \\ *[8pt] 
                H=-2 (\Phi-1)^2 & k^{2}>2.  
\end{array}  
\right.  
\end{equation}

Finally, substituting (31) in Eqs.(26)-(28) we get the dynamical equations  
for the different ranges of the parameter $k$
\vskip 1cm

$ k^2<2 $:
\nopagebreak[2]
\begin{eqnarray}
\beta' & = & \left({1\over 2}\sqrt{1-b^2}-{3 \over 4b}\beta\right)L
        -{2b^2+1\over 2b}\beta \Gamma^2 \nonumber \\
\Phi' & = & {1\over 4b}\left[\left(3-3\Phi-2b^2\right)L
         +2\left[1-\left(2b^2+1\right)\Phi\right]\Gamma^2\right] \\
\Gamma' & = & {\Gamma \over 4b}\left[-3L+2\left(2b^2+1\right)
\left(1-\Gamma^2\right)-2\Phi\right], \nonumber 
\end{eqnarray}
\vskip 1cm

$k^2=2 $:
\begin{eqnarray}
 \beta' & = & {1 \over 2}\left(1-4 \beta\right) L-2 \beta \Gamma^2  
 \nonumber \\
 \Phi' & = & {1 \over 2}\left(3-4 \Phi\right) L+\left(1-2 \Phi\right) 
 \Gamma^2 \\
 \Gamma' & = & -\Gamma \left( 2L- {1 \over 2}+2 \Gamma^2\right), 
 \nonumber
\end{eqnarray}
\vskip 1cm

$k^2>2$:

\begin{eqnarray}
\beta' & = &  \left( {1 \over 2}\sqrt{1+b^2}-{3+4b^2 \over 4b} \beta \right) L
       -{1 \over 2b} \beta \,\Gamma^2\left( 1+2b^2\right) \nonumber \\
\Phi' & = & {1 \over 4b} \left( \left[ \left( 3+4b^2\right)\left( 1-
\Phi\right)  -2b^2\right] L +2\left[ 1-\left( 1+2b^2\right)\Phi\right]
\Gamma^2 \right) \\
\Gamma' & = & -{\Gamma \over 4b}\left[ \left( 3+4b^2\right) L 
-2\left( 1-2b^2\right) +2\left( 1-4b^2\right)\Phi
+2\left( 1+2b^2\right)\Gamma^2\right]. \nonumber
\end{eqnarray}
The prime stands for the derivative with respect to the new time 
$\tau$ and $L$ is defined in (39).

From (39)-(40) we see that there are three different invariant
subspaces, one for each value of the parameter $\epsilon$. The first is
the closed surface $L=0$ which corresponds to the linear case
$\epsilon=0$. The other two subspaces are separated by the surface
$L=0$. The invariant subspace for $\epsilon=1$ is a compact region
limited by the surfaces $L=0$ and $\Gamma=0$ and, finally, the invariant
subspace for $\epsilon =-1$ is the unbounded region outside $L=0$. 

The most satisfactory scenario to give the explicit form of the
asymptotic behaviour of the models is the one in which we have variables
which are bounded for all times. This is guaranteed for both
$\epsilon=0$ and $\epsilon=1$ since we have seen that in these two cases
$\dot\tau$ remains always positive and the phase-space is bounded. As
explained above this is not the case for $\epsilon=-1$. We will see from
the numerical calculation of the trajectories that, as expected, for
$\epsilon=-1$ some trajectories go to infinity in a finite time.
Probably one can find a different set of variables for which this part
of the phase space becomes bounded, although it is unlikely that this
can be done for the whole phase-space.

\section{The equilibrium points and the associated  metrics}  
 
We are interested in looking for results concerning no-hair theorems, thus 
for each of these dynamical systems we will look for the equilibrium 
points and their stability. This  
is important since attractors in phase-space will be related to the 
behaviour of the models in the limit  
$\tau \rightarrow \infty$ ($t \rightarrow \infty$) while sources, the 
unstable fixed points, will be related  
to the behaviour near the initial singularity $\tau \rightarrow - \infty$ 
($t \rightarrow 0$).  

Before giving the explicit expressions of the equilibriun points of the
system, it is interesting to investigate whether the equilibrium points
represent self-similar solutions. It has been shown that the homogeneous
but anisotropic models (Bianchi models) evolve towards self-similar
solutions \cite{w2} and it has been conjectured that the same behaviour
is shared by the $G_2$ metrics \cite{w1}. We will prove that the
solutions considered in this paper do approach self-similar solutions.
Here is a sketch of the proof of this statement. 

When $\epsilon =0,1$ the equilibrium points are characterized by finite
constant values of the variables $\beta$, $\Phi$ and $\Gamma$, then for
an equilibriun point equation (33) takes the form: \begin{equation}
{\ddot T \over T}=(1-W) \left( {\dot T \over T} \right)^2 \end{equation}
where $W$ is a function both of the constant values of $\beta$, $\Phi$
and $\Gamma$ and the parameters of the model. This equation has only two
types of solutions depending on whether or not $W$ is zero. When $W \neq
0$ the solution is of the form $T=t^{1/W}$ while when $W=0$ it is an
exponential, that is, $T=e^{{\overline W}t}$, where ${\overline W}$ is
another constant. For the power law solution $T=t^{1/W}$, equations (34)
and (39) force $\epsilon=0$ which in turn forces $Z(z)$ to be linear in
$z$. 
  
Writing down the equations for a homothetic vector $\vec\xi$,  
\begin{equation}
{\pounds_{\vec\xi}} g_{ab}=2g_{ab},
\end{equation}
we  
find that for both solutions, exponential and power law, there is at 
least one homothetic vector with components 
$\xi^{a}=(a_{0},a_{1}x,a_{2}y,0)$ in the exponential case and  
$\xi^{a}=(b_{0}t,b_{1}x,b_{2}y,b_{0}z)$ in the power law case, where 
both $a_{m}$ and $b_{m}$ are  
constants (the details of this calculation are given in the Appendix)  

To give a qualitative behaviour of the solutions of the system (41)-(43)
it is important to obtain the equilibrium points and their local
stability. Let us first note that the system admits the symmetry
$\Gamma\to -\Gamma$. Therefore, the study of the equilibrium points will
be restricted to $\Gamma\geq 0$. They can be found explicitely and are
given by:

\subsection*{$P_{1}$ and $P_{2}$}  

\begin{displaymath}
\begin{array}{lll}
\beta^2+\Phi^2=1\; & \Gamma=0\; & k^2<2 \\
\beta^2+\Phi^2-\Phi=0\; & \Gamma=0\; &k^2=2 \\
\beta^2+\left( 1-\Phi\right)\left(1-3\Phi\right)=0\; & \Gamma=0\; & k^2>2
\end{array}
\end{displaymath}

These points lie on the surface $L=0$ corresponding, therefore, to
$\epsilon=0$. All of them are the sources of the entire phase-space,
i.e: the attractors at early times. This set of points reduces to two
separate points when the constraint equation (38) is taken into account.
These are the two points given by the intersection in $\Gamma=0$ of the
surface $L=0$ and the plane defined by the first constraint equation
(38). For $k^{2} \geq 6$ one of them becomes a saddle, attracting the
trajectories lying on $L=0$. Since $\Gamma=0$, the corresponding metrics
describe massless scalar field cosmologies. The line element and the
scalar field are:
\begin{eqnarray}
ds^{2} & = & Dt^{C_{1}}E_{z}^{C_{\pm}}(-dt^{2}+dz^{2})
        +t^{1+C_{2}}E_{z}^{1+n} dx^{2} 
         + t^{1-C_{2}}E_{z}^{1-n} dy^{2} \nonumber \\  
\phi & = & \left( \pm C_{2}\frac{\sqrt{k^2+2-2n^2}}{2n}-\frac{k}{2} \right) 
           \log t  
     -\frac{C_{\pm}}{k} \log E_{z}  
 \end{eqnarray}
where 
\begin{eqnarray} 
& E_{z}=Az+B & \\
& C_{1}=1+\frac{C_{2}}{n} (1+C_{\pm}) \nonumber &  \\  
& C_{2}=n \left( 1 \pm \sqrt{\frac{8}{k^2+2}} \right). & \nonumber  
\end{eqnarray}  
The $-$ and $+$ signs appearing in the definition of $C_{2}$ 
correspond to $P_{1}$ and $P_{2}$  
respectively.  
When $n ^2=1$ and $mn=k/2$  $C_{\pm}=0$ and it is easy to see by a simple
coordinate transformation that (46) is a LRS Bianchi type I solution.

\subsection*{$P_{3}$}  
  
\begin{displaymath}
\begin{array}{llll}
\beta=\frac{2b \sqrt{1-b^2}}{3}\; & \Phi=1-{2\over 3}b^2 \;
 & \Gamma=0 \; & k^2<2 \\
\beta={1 \over 4} \; & \Phi={3 \over 4} \; & \Gamma=0 \; & k^2=2 \\
\beta={2b \sqrt{1+b^2} \over 3+4b^2} \; & \Phi={3+2b^2 \over 3+4b^2} \;
 & \Gamma=0 \; & k^2>2 
\end{array}
\end{displaymath}

This point corresponds, again, to a massless scalar field but wiht
$\epsilon=1$. It is a saddle for all $k$, attracting the trajectories
with $\epsilon=1$ in $\Gamma=0$. The metric and the scalar field
corresponding to this point are:
\begin{eqnarray}    
ds^{2} & = & De^{(2+C_\pm)at}E_{z}^{C_{\pm}} (-dt^{2}+dz^{2})  
       + \left( e^{at}E_{z} \right) ^{1+n} dx^{2}  
       + \left( e^{at}E_{z} \right) ^{1-n} dy^{2} \nonumber \\    
\phi & = & -\frac{C_{\pm}}{k} \left( at+ \log E_{z} \right)
\end{eqnarray}
where  
\begin{displaymath}
 E_{z}=A \cosh az +B \sinh az 
\end{displaymath}  
and $A$ and $B$ are arbitrary constants.  
  
By making the change: $T=e^{at}E_z$, $Z=e^{at}E_z'$ when $A>B$ we see
that (47) is an anisotropic Bianchi type I space time. (For $A \leq B$
the model has spacelike curvature singularities). In the special case
when $n^2=1$ and $mn=k/2$ this turns out to be Minkowski flat metric. 
  
\subsection*{$P_{4}$}  
\begin{displaymath}
\begin{array}{llll}
\beta=0 \; & \Phi={1 \over 1+2b^2} \; &
\Gamma={2b \sqrt{1+b^2} \over 1+2b^2}\; & k^2<2 \\
\beta=0 \; & \Phi={1 \over 1+2b^2} \;
& \Gamma={2b \sqrt{1-b^2} \over 1+2b^2} \; & k^2>2
\end{array}
\end{displaymath}  

For this point $\epsilon=0$. It does not exist for $k^{2}=2$ or
$k^{2} \geq 6$. For $k^{2}<2$ this is the attractor for trajectories with 
$\epsilon=0$ and $\epsilon=1$  
(also for some of the trajectories with   
$\epsilon=-1$).   
For $2<k^{2}<6$ it is a saddle, attracting only the trajectory with   
$\epsilon=0$. In this case:  
\begin{eqnarray}    
ds^{2} & = & \left[ {k^{2}-2 \over 2 \sqrt{6-k^{2}}} t 
\right]^{\frac{4}{k^{2}-2}}  
          \left[ \frac{E_{z}^{C_{\pm}}}{2 \Lambda}(-dt^{2}+dz^{2})  
                  + E_{z}^{1+n} dx^{2} + E_{z}^{1-n} dy^{2} \right] 
           \nonumber \\  
\phi & = & \frac{2k}{2-k^2} \log \left[ {k^{2}-2 \over 2 \sqrt{6-k^{2}}} t 
\right]
-\frac{C_{\pm}}{k} \log E_{z},   \;\;\;\;\;   E_{z}=Az+B.   
\end{eqnarray}  
When $n^2=1$ and $mn=k/2$ ($C_{\pm}=0$) it is easy to see that this is a
flat FRW space time.

\subsection*{$P_{5}$}  
\begin{displaymath}
\begin{array}{llll}
\beta={2b^3 \over \sqrt{1-b^2}} \; & \Phi=1-2b^2 \;
& \Gamma={2b \over \sqrt{1-b^2}} \; & k^2<2 \\
\beta=0 \; & \Phi={1 \over 2} \; & \Gamma={1 \over 2} \; & k^2=2 \\
\beta={2b^3 \sqrt{1+b^2} \over 1+5b^2+4b^{4}} \; & \Phi={1+2b^2 \over 1+4b^2}
\; & \Gamma={2b \sqrt{1+b^2} \over 1+5b^2+4b^{4}} \; & k^2>2 
\end{array}
\end{displaymath}

The solution corresponding to this point has $\epsilon=-1$ when
$k^{2}<2$, $\epsilon=0$ when $k^{2}=2$ and $\epsilon=1$ when $k^{2}>2$.
It is a saddle for $k^{2} \leq 2$ and a sink for $k^{2}>2$. When
$k^{2}>2$ this is the attractor for the trajectories with $\epsilon=1$
and when $k^2=2$ is the attractor for both trajectories with
$\epsilon=0$ and $\epsilon=1$. The metric and the scalar field are: 
\begin{eqnarray}  
ds^{2} & = & \frac{A^{2}-\epsilon a^{2}}{2 \Lambda} e^{C_{1}t} 
E_{z}^{C_{\pm}}  
            (-dt^{2} +dz^{2}) \nonumber \\  
       &  &  \mbox{} +e^{[(A^{2}+n \epsilon a^{2})/A]t} E_{z}^{1+n} dx^{2}  
             +e^{[(A^{2}-n \epsilon a^{2})/A]t} E_{z}^{1-n} dy^{2} 
             \nonumber \\  
\phi & = & -\frac{1}{k} (C_{1} t+ C_{\pm}\log E_{z})   
\end{eqnarray}    
where
\begin{displaymath}
C_{1}=\frac{k^{2}A^{2}\mp\epsilon a^{2} k \sqrt{k^{2}+2-2n^{2}}}{2A}, \quad
a^{2}=\epsilon \frac{k^{2}-2}{k^{2}+2} A^{2} 
\end{displaymath}
\begin{displaymath}
E_{z}=\left\{ \begin{array}{ll}
            \bar{A} \cosh (az)+\bar{B} \sinh (az)  &  \epsilon=1  \\  
            \bar{A} z+\bar{B}  &  \epsilon=0  \\  
            \bar{A} \cos (az)+ \bar{B} \sin (az) &  \epsilon=-1  
            \end{array}  
      \right.      
\end{displaymath}  
where $A$, $\bar{A}$ and $\bar{B}$ are arbitrary constants.  
  
This solution was previously found in \cite{ajp}. As in the previous
cases, when $n^2=1$ and $mn=k/2$ it is homogeneous: when $k^2<2$, it is
a Kantowski-Sachs space time; when $k^2=2$ it is a flat FRW model while
for $k^2 >2$ it is a LRS Bianchi type III solution. This last case can
be seen by making the coordinate transformation: 
\begin{displaymath}
X=-dx-\ln E_z,\;\;\;\; Z=-\frac{1}{d^2}\frac{E_z'}{E_z}e^{-dx},
\end{displaymath}
where $d$ is a constant related with the constants appearing in the solution.  

To complete the description of the evolution of the solutions we have
solved the system (41)-(43) with the constraint equation (38) by
numerical integration. The phase-space diagrams so obtained are plotted
in Fig.1 for different ranges of the parameter $k$. The bold line
represents the invariant subspace $L=0$. This line splits the
phase-space into two separate invariant subspaces: the inner region
correponds to $\epsilon=1$ ($L>0$) and the outer region to $\epsilon=-1$
($L<0$). The two fixed points $P_{1}$ and $P_{2}$ are located in which
constitutes the common points of the two invariant subspaces defined by
$L=0$ and $\Gamma=0$ (subspace of massless scalar field solutions).
$P_{3}$, lies in the invariant subspace $\Gamma=0$, $L>0$. $P_{4}$ lies
in the line $L=0$ and exists only when $k^2<2$ and when $2<k^2<6$. It
aproaches the point $P_1$ as $k^2\to 6$ and the point $P_5$ as $k^2\to
2$. Finally, $P_{5}$ does not stay in the same invariant subspace for
all $k$. As a matter of fact, this point is in the region defined by
$L<0$ when $k^2<2$, is on $L=0$ when $k^2=2$, and lies in $L>0$ when
$k^2>2$.

\section{Conclusions}  
  
In this work we have considered a class of inhomogeneous solutions with
a minimally coupled scalar field with an exponential potential. This
class is more general than that studied in paper I. Therefore, we can
analyse the effect produced by complex initial inhomogeneities on the
dynamics of scalar field cosmologies and specifically on the
isotropization of the models. 

When the inhomogeneity was linear, as that described in paper I, or when
the solutions were homogeneous (Bianchi type), the dynamics of the
solutions was completely determined by the value of the parameter $k$
and it was shown that for $k^2 \leq 2$ all the solutions reached
homogeneity and isotropy but when $k^2>2$ while all the models still
attained homogeneity only a subclass of measure zero reached isotropy. 

We have found the equilibrium points of the dynamical systems associated
with part of the solutions (23) and we have shown that all of them are
self-similar. Except the point $P_3$, they correspond to inhomogeneous
space times with a scalar field configuration that cannot be related to
a perfect fluid source. However, in the special case when the scalar
field is homogeneous ($n^2=1$ and $mn=k/2$), they are all homogeneous
perfect fluid space times. The fixed point $P_{3}$ represents a
homogeneous perfect fluid space time for every value of the parameters
defining the particular model. 

We can conclude that although the dynamics is still determined by $k$,
as in the former cases, the result refering to the isotropization or
homogeneization of the space time is completely different. Actually,
when $k^2<2$ solutions with $\epsilon=0,1$ evolve towards the point
$P_4$ which is inhomogeneous. Although the point $P_3$ is an attractor
for a subset of solutions, it is a saddle point, attracting only the
massless scalar field solutions. Hence, the inhomogeneities prevent in
this case the solutions from isotropizing. Only in the particular case
when $n^2=1$ and $mn=k/2$ (corresponding to a homogeneous scalar field)
we do have a similar situation to the previous cases. When $k^2\geq 2$
the point $P_5$ becomes the attractor for the solutions. 

In the region where $\epsilon=-1$ the situation is quite different since
the variables used are not bounded. If $k^2<2$ the stable manifold of
the fixed point $P_{5}$ splits this region into two separate regions.
The solutions with trajectories taking place in the inner region are
attracted by the inhomogeneous fixed point $P_4$, whereas those on the
stable manifold approach the inhomogeneous solution $P_{5}$. Those with
trajectories in the unbounded outer region escape to infinity. When $k^2
\geq 2$ this separation does no longer exist and all the solutions with
($\epsilon=-1$) escape to infinity. 

Points $P_1$ and $P_2$ describe the solutions ``close'' to the initial
singularity $t=0$. On both points the potential vanishes since near the
singularity the kinematical term is the leading one in the stress-energy
tensor. By making a coordinate transformation the line element and the
scalar field can be cast into the form:
\begin{eqnarray}
ds^2 & = & -a_0(z)dt^2+a_1(z)t^{2p_1}dx^2+a_2(z)t^{2p_2}dy^2
         +a_3(z)t^{2p_3}dz^2,\nonumber \\
\phi & = & q\ln t+b(z),\nonumber 
\end{eqnarray}
where $p_i$ and $q$ are constants verifying
\begin{displaymath}
p_1+p_2+p_3=p_1^2+p_2^2+p_3^2+q^2=1. 
\end{displaymath}
This agrees with \cite{bl} in the sense that the above metric could
represent the asymptotic form of a ``general'' scalar field cosmological
solution near the singularity. 

It is important to note that solutions corresponding to a cosmological
constant can be analysed by letting $k=m=0$. In this case all the
solutions evolve to homogeneous space times. 

As we mentioned in section 2, in order not to extend this paper the
general exponential case will be considered in a separate paper. It is
relevant just to note that the asymptotic behaviour in that case is
similar to that found here in the sense that even when $k^2<2$ the
equilibrium points of the system are inhomogeneous. 

\vskip 1cm 
\noindent
{\bf Acknowledgments}

We are grateful to Prof. A.A.Coley and Dr. R.van den Hoogen for
stimulating discussions. We are indebted to Prof. A.Feinstein for his
comments. This work is supported by a grant DGICYT PB93-0507. I.O.'s
work is supported by a fellowship from the DGICYT FP94.

\vskip 2cm
\appendix
\section{Appendix}
\setcounter{equation}{0}
\renewcommand{\theequation}{A.\arabic{equation}}

In this appendix we are going to show that all the fixed points of the
system under study are indeed self-similar solutions. By inspecting the
equilibrium points listed in Section 3 it is easy to see that for the
fixed points the functions $e^{f(t)}$ and $Q(t)$ are powers of the
function $T(t)$:
\begin{equation}
e^{f(t)}=T(t)^{C_{1}} \mbox{\hspace{2cm}}  Q(t)=T(t)^{C_{2}},
\end{equation}
where $C_1$ and $C_2$ are constants depending on the equilibrium point
we are considering.

The equations for the existence of a homothetic vector $\vec{\xi}$,
taking into account Eq.(A.1), reduce to the following equations:

\begin{eqnarray}
C_{1}{\dot{T} \over T}\xi^{0}+ 2 \alpha {Z' \over Z} \xi^{3}+2
\partial_{0} \xi^{0}=2  && \\
(1+ C_{2}) {\dot{T} \over T}\xi^{0}+(1+n) {Z' \over Z} \xi^{3}+2
\partial_{1} \xi^{1}=2  && \\
(1- C_{2}) {\dot{T} \over T}\xi^{0}+(1-n) {Z' \over Z} \xi^{3}+2
\partial_{2} \xi^{2}=2  && \\
\partial_{0} \xi^{0}=\partial_{3} \xi^{3} && \\
DT^{C_{1}}Z^{2 \alpha} \partial_{1} \xi^{0}-T^{1+C_{2}}Z^{1+n}
\partial_{0} \xi^{1}=0 && \\
DT^{C_{1}}Z^{2 \alpha} \partial_{2} \xi^{0}-T^{1-C_{2}}Z^{1-n}
\partial_{0} \xi^{2}=0 && \\
\partial_{3} \xi^{0}-\partial_{0} \xi^{3}=0 && \\
T^{2C_{2}}Z^{2 n} \partial_{2} \xi^{1}+ \partial_{1} \xi^{2}=0 && \\
DT^{C_{1}}Z^{2 \alpha} \partial_{1} \xi^{3}+T^{1+C_{2}}Z^{1+n}
\partial_{3} \xi^{1}=0 && \\
DT^{C_{1}}Z^{2 \alpha} \partial_{2} \xi^{3}+T^{1-C_{2}}Z^{1-n}
\partial_{3} \xi^{2}=0
\end{eqnarray}

We have shown that the function $T(t)$ in the equilibrium points
verifies the Eq.(44) giving two different solutions:

I)  $W=0  \Rightarrow  { \dot{T} \over T}=k=constant$.

Looking at the equations (A.2)-(A.11) one can try the following solution:

\begin{equation}
\xi^{0}=constant \;\; \xi^{1}=\xi^{1}(x) \;\;
\xi^{2}=\xi^{2}(y) \;\; \xi^{3}=0. 
\end{equation}

Equations (A.5)-(A.11) are automatically satisfied and equations (A.2)-(A.4)
determine completely $\vec{\xi}$:

\begin{equation}
\vec{\xi}=\left\{ {2 \over C_{1}k},{C_{1}-C_{2}-1 \over C_{1}}x,
{C_{1}+C_{2}-1 \over C_{1}}y,0\right\}.
\end{equation}

II)  $W  \neq 0 \Rightarrow { \dot{T} \over T}={1 \over Wt}\; \mbox{and}\;
{Z' \over Z}={1 \over z}$

In this case equations (A.2) and (A.5) suggest a solution of the form:

\begin{equation}
\xi^{0}=ct \;\; \xi^{1}=\xi^{1}(x) \;\;
\xi^{2}=\xi^{2}(y) \;\; \xi^{3}=cz, 
\end{equation}
where $c$ is a constant to be determined.

As before, the actual form is completely determined by equations
(A.2)-(A.4), the remaining equations being automatically satisfied. Then
we obtain:

\begin{eqnarray}
\vec{\xi}& = &  {W \over C_{1}+W(C_{3}+2) }\left\{ 2t, \left( {C_{1}-C_{2}-1
\over W} +1+C_3-n \right) x, \right. \nonumber \\
 & & \mbox{} \left. \left( {C_{1}+C_{2}-1 \over W}+1
 +C_{3}+n \right)y,2z \right\}
\end{eqnarray}

\newpage
\thispagestyle{empty}
\section*{Figure Caption}

Figure 1: Phase-space diagrams are drawn for (a) $k^2=1$, (b) $k^2=2$,
\linebreak (c) $k^2=4$ and (d) $k^2=10$. Each diagram represents the
typical form of the phase-space for the four significant ranges of the
parameter k. $\Gamma$ is shown on the vertical axis and $\beta$ on the
horizontal axis. The bold line stands for the separatrix $L=0$, the
arrows show the direction of increasing time, and the dots represent the
fixed points, which are named as in the body of the paper. We can easily
distinguish the four important invariant subspaces: $\Gamma=0$ (massless
scalar field), $\epsilon=-1$ (the unbounded region outside $L=0$),
$\epsilon=0$ (the separatrix $L=0$ itself) and $\epsilon=1$ (the bounded
region iside $L=0$).

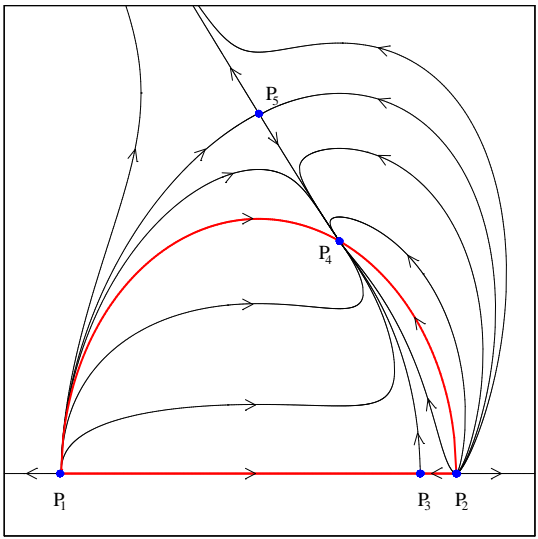
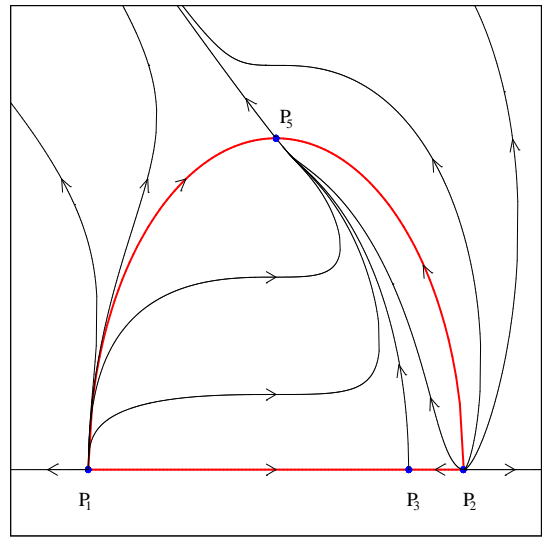
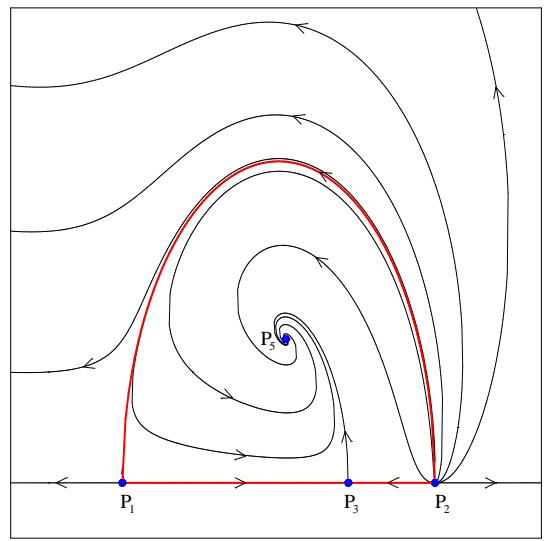
\end{document}